\documentclass[12pt,preprint]{aastex}
\newcommand{\RCB}{$\rho$ CrB}

\newcommand{\MJ}{$M_{J}$}

\newcommand{\degrm}{^{\circ}}
\newcommand{\Mo}{$M_{\odot}$}
\newcommand{\mas}{{$mas$}}
\newcommand{\GHB}{Gatewood, Han \& Black (2001)}
\newcommand{\HBG}{Han, Black \& Gatewood (2001)}
\newcommand{\GHBs}{Gatewood et al. (2001)}
\newcommand{\HBGs}{Han et al. (2001)}

\slugcomment{July 19th, 2001}

\shorttitle{Analysis of the Hipparcos Observations of extrasolar
  planets}
\shortauthors{Zucker \& Mazeh}

\begin{document}

\title{Analysis of the Hipparcos Observations\\
 of the Extrasolar Planets and the Brown-Dwarf Candidates}

\author{Shay Zucker and Tsevi Mazeh}
\affil{School of Physics and Astronomy, Raymond and Beverly Sackler
Faculty of Exact Sciences, Tel Aviv University, Tel Aviv, Israel}
\email{shay@wise.tau.ac.il, mazeh@wise.tau.ac.il} 

\affil{accepted for publication by the Astrophysical Journal}

\begin{abstract}

We analyzed the Hipparcos astrometric observations of 47 stars that were
discovered to harbor giant planets and 14 stars with brown-dwarf
secondary candidates. The Hipparcos measurements were used {\it
together} with the corresponding stellar radial-velocity data to derive
an astrometric orbit for each system. To find out the significance of
the derived astrometric orbits we applied a ``permutation'' technique by
which we analyzed the {\it permuted} Hipparcos data to get false
orbits. The size distribution of these false orbits indicated the range
of possibly random, false orbits that could be derived from the true
data. These tests could not find any astrometric orbit of the planet
candidates with significance higher than 99\%, suggesting that most if
not all orbits are not real. Instead, we used the Hipparcos data to set
upper limits on the masses of the planet candidates. The lowest derived
upper limit is that of 47~UMa --- 0.014 \Mo, which confirms the
planetary nature of its unseen companion. For 13 other planet candidates
the upper limits exclude the stellar nature of their companions,
although brown-dwarf secondaries are still an option. These negate the
idea that all or most of the extrasolar planets are disguised stellar
secondaries. Of the 14 brown-dwarf candidates, our analysis reproduced
the results of Halbwachs et al.\, who derived significant
astrometric orbits for 6 systems which imply secondaries with stellar
masses. We show that another star, HD~164427, which was discovered only
very recently, also has a secondary with stellar mass. Our findings
support Halbwachs et al.\ conclusion about the possible existence
of the ``brown-dwarf desert'' which separates the planets and the
stellar secondaries.

\end{abstract}


\section{Introduction}

About fifty candidates for extrasolar planets have been announced over
the past five years (e.g., Schneider 2001).  In each case, precise
stellar radial-velocity measurements indicated the presence of a
low-mass unseen companion, with a minimum mass between 1 and about 10
Jupiter masses (\MJ). The actual masses of the companions are not known,
because the inclination angles of their orbital planes cannot be derived
from the spectroscopic data.

Precise astrometry of the orbit, from which we can derive the
inclination, can yield the secondary mass, at least for the cases where
the primary mass can be estimated from its spectral type. The Hipparcos
accurate data, which have already yielded numerous orbits with small
semi-major axes (ESA 1997; S\"oderhjelm 1999) down to a few
milli-arc-sec ($mas$), could in principle be used for this purpose. The
satellite data could be used either to detect small astrometric orbits
of the stars that are known to host planet candidates, or to put upper
limits to the size of the stellar astrometric motions, as was done by
Perryman et al.\ (1996) for the first three planets discovered.

In principle, the analysis of the Hipparcos data of the stars hosting
planet candidates {\it together} with their precise radial-velocity
measurements can be a powerful technique. The idea is that the
combination of the Hipparcos and the radial-velocity data can reveal
small astrometric orbits which could not have been seen with the
astrometric data alone. Mazeh et al.\ (1999) and Zucker \& Mazeh (2000)
were the first to apply this technique, followed by Halbwachs
et al.\ (2000), \GHB\ and \HBG.

However, such analyses, including our own (Mazeh et al.\ 1999 and Zucker \& Mazeh
2000), can be misleading.  
As has been shown by
Halbwachs et al.\ (2000), one can derive a small {\it false} orbit with the size
of the typical positional error of Hipparcos --- about 1 \mas, caused by the scatter
of the individual measurements.  The astrometric orbit of \RCB\ derived by
\GHBs\ and few of the astrometric orbits derived by \HBGs\ were suspicious
in particular, because the inclination angles implied by those
astrometric orbits were very small. \HBGs\ found eight out of 30 systems
with an inclination smaller or equal to $0\fdg5$, four of which they
categorized as highly significant. The probability of finding such small
inclinations in a sample of orbits that are {\it isotropically} oriented
in space is extremely small, indicating either a problematic derivation
of the astrometric orbit, or, as suggested by \HBGs, some serious
orientation bias in the distribution of the inclination angles.

Indeed, two very recent independent studies of the statistical
significance of the \RCB\ orbit found by \GHBs\ indicated that the
finding is not highly significant. Pourbaix (2001), who studied the
significance of the derived Hipparcos orbits of 42 stars that host
planet candidates, found that the statistical significance of the \RCB\
orbit is at the 99\% level. He used an F-test to evaluate the
improvement of the fit to the Hipparcos data resulting from the
additional parameters of the binary orbit. He also concluded that the
other orbits of \HBGs\ are statistically non significant. The use of the
F-distribution assumes Gaussianity of the individual measurements, an
assumption that might not be well justified for the Hipparcos data.
Arenou et al.\ (1995) proved that the Hipparcos parallaxes and
zero-points are Gaussian, but they do not analyze the distribution of
the individual measurements of each star.  We (Zucker \& Mazeh 2001)
avoided the assumption of Gaussianity by using the permutation test,
which belongs to the class of distribution-free tests (e.g. Good 1994)
and thus is more robust against modeling problems of the measurement
process.

We performed the permutation test by generating simulated data from the
very same astrometric measurements of \RCB. If there was some evidence of
an orbit in the measurements, it should be ruined by the permutation,
and thus no random permutations would yield a comparable
orbit. However, if the derived orbit was only spurious, some random
permutations should be able to reproduce a similar effect. In a sense,
we let the data ``speak for themselves'' and do not have to assume any
specific distribution for the measurements or the errors. We found that
the significance of the astrometric solution of \RCB\ is at the
97.7\% level. 

The study of the significance of the \RCB\ astrometric detection led
us to check carefully {\it all} the announced extrasolar planet and
brown dwarf candidates with available data. All together we present
here an analysis of 47 planet candidates and 14 brown dwarfs. We first
derive from the Hipparcos data the best astrometric orbit and then
test unequivocally its significance, with an approach which is free of
any assumptions about the errors of the Hipparcos measurements.  All
the derived orbits of the stars that harbor extrasolar planets turned
out to be insignificant. Pourbaix and Arenou (2001) reached similar
conclusions. Although the derived orbits are insignificant, we
nevertheless used them to derive upper limits for the corresponding
astrometric motions and for the masses of the unseen companions.

The previous study which derived upper limits for planet candidates
(Perryman et al.\ 1996) searched the Hipparcos data for any possible
astrometric orbital periodicity for each of the three stars considered
then. They derived a periodgram from the corresponding astrometric
amplitudes and obtained an upper limit for the stellar astrometric
motion from the amplitude corresponding to the radial-velocity
period. 
That algorithm does not utilize all the known radial-velocity elements.
We, on the other hand, derive astrometric upper limits by
combining the Hipparcos data {\it together} with all the radial-velocity
information available, as we did in our previous work.

Section 2 presents our analysis, Section 3 the results for the whole
sample and Section 4 concentrates on HD~164427, another star that hosts
brown-dwarf candidate that is probably a massive companion. Section 5
discusses our finding.

\section{Analysis}

\subsection{Orbital Solution}

For each of the extrasolar planets and the brown-dwarf candidates we
first derived the best astrometric orbit from the Hipparcos data,
following closely our analysis of \RCB\ (Zucker \& Mazeh 2001). The
analysis assumed that the spectroscopic and astrometric solutions have
in common the following elements: the period, $P$; the time of
periastron passage, $T_0$; the eccentricity, $e$; the longitude of the
periastron, $\omega$.  In addition, the spectroscopic elements include
the radial-velocity amplitude, $K$, and the center-of-mass radial
velocity $\gamma$.  We have three additional astrometric elements ---
the angular semi-major axis of the photocenter, $a_0$; the inclination,
$i$; the longitude of the nodes, $\Omega$. In addition, the astrometric
solution includes the five regular astrometric parameters --- the
parallax, the position (in right ascension and declination) and the
proper motion (in right ascension and declination). All together we had
a 14-parameter model to fit to the spectroscopic and astrometric data.

For some extrasolar planets the discoverers made the individual stellar
radial velocities available. 
Unfortunately, for many planets the individual
velocities on which the discovery was based were not available to us.
In such cases we used the published elements and their errors, as independent
measurements of the elements, instead of the unavailable radial
velocities.

The 14 elements are not all independent. From $K$, $P$ and
$e$ we can derive the projected semi-major axis of the primary orbit ---
$a_{1,phys}\times \sin i$, in physical units. This element, together
with the inclination $i$ and the parallax, yields the angular semi-major
axis of the primary, $a_1$.  {\it Assuming the secondary contribution to
the total light of the system is negligible}, this is equal to the
observed $a_0$.

\subsection{Statistical Significance}

As pointed out by Halbwachs et al.\ (2000), the scatter of the actual
measurements can cause a false ``detection'' of a very small semi-major
axis, even without any real astrometric motion.  To find the statistical
significance of the derived astrometric orbit in each case we followed
the permutation test (e.g. Good 1994) that we applied to \RCB\ (Zucker
\& Mazeh 2001). For each star we generated simulated permuted
astrometric positions by using the IAD Hipparcos measurements (ESA 1997)
and permuting the actual timing of the observations, modifying the
partial derivatives with respect to the five astrometric parameters (ESA
1997) accordingly.  We then analyzed the permuted astrometric positions
together with the actual radial-velocity data, deriving a new false
astrometric orbit. This resampling process was repeated 1000 times for
each star.

For most of the original measurements there are two stellar positions,
one derived from the NDAC and the other from the FAST consortia. The
two positions are, obviously, not independent, but have an assigned
correlation.  In our permutation we kept the pairing of
the corresponding NDAC and FAST positions, while permuting the timings
among the pairs.  
Thus, we actually applied separate permutations to the set of
single measurements, where only one consortium produced a result, and
to the set of paired permutations where there are two measurements for
each timing. The partial derivatives of the residuals with respect to
the basic astrometric parameters, which are part of the original IAD, had to
be dealt with very carefully, and we had to re-compute them using the
new timings (ESA 1997).
In an earlier study this was not done correctly and it
caused us to falsely assign a higher significance to the astrometric
orbit of \RCB. This emphasizes the importance of
the careful calculation of the partial derivatives.

The distribution of the ensemble of falsely detected semi-major axes
indicated the range of possible random detections. For example,
$a_{99}$---the 99-th percentile, denotes the semi-major axis size
for which 99\% of the simulations yielded smaller
values. Consequently, an astrometric orbit is detected with a
significance of 99\% if and only if the corresponding semi-major axis
is larger than $a_{99}$.

As an illustration, Figure~1a shows the histogram of the semi-major
axis derived using random permutations of the Hipparcos data of
HD~209458. This star's inclination is known to be close to $90 \degrm$
through the combination of radial velocities and transit measurements
(Charbonneau et al.\ 2000; Henry et al.\ 2000; Brown et al.\ 2001).
The Hipparcos derived semi-major axis is 1.76 \mas, which is marked in
the figure by an arrow. One can clearly see that many random
permutations led to larger semi-major axes, a fact that renders
this derived value insignificant. The derived value is obviously false
since the known inclination implies a value of less than a
micro-arc-second. 

The opposite case --- of a significant derived orbit of HD~164427, is
shown in Figure 1b and is discussed in Section 4.

\subsection{Astrometric Upper Limits}

It turned out that our stringent threshold --- $a_{99}$, rendered most
of the astrometric orbits derived from the Hipparcos data insignificant.
We tried instead to use the Hipparcos data for deriving an upper limit
to the astrometric motion (Perryman et al.\ 1996). Note that $a_{99}$
can {\it not} be used as an upper limit to the stellar astrometric
orbit, even if the derived semi-major axis, $a_{derived}$, is smaller
than $a_{99}$. We have no argument to negate the possibility that
the true semi-major axis, $a_{true}$, is
$$ a_{derived} \leq a_{99} < a_{true} \ . $$
In fact, $a_{99}$ was computed {\it a priori} with the assumption that
there is no astrometric orbit, and therefore denotes the range in which
$a_{derived}$ could be found if $a_{true}=0$. The fact that
$a_{derived}$ was found within this range only tells us that
$a_{true}=0$, or very close, is a possibility, but does not constrain its permitted
values.

In order to set a reliable upper limit to the astrometric orbit, we have
to consider the range of permitted values of $a_{true}$, given the fact
that the derived value is $a_{derived}$. We could run simulations that
turn this question around and find {\it a priori} the range of
$a_{derived}$ for every possible value of $a_{true}$, as we actually did
for the case $a_{true}=0$.  However, such an approach would be
prohibitively extensive in numerical computations. Instead we adopt the
usual approach and used the error estimate of $a_{derived}$, calculated
from the second derivatives of the $\chi^2$ function at
$a_{derived}$. We are assuming that the effects of non-normality and
skewness of the error distribution are weak at $a_{derived}$. This
assumption is obviously wrong at $a = 0$, since we expect only positive values to
be derived, but is more probable at a non-vanishing value. We therefore
use the 2.3$\sigma$ confidence interval around $a_{derived}$ to set an
upper limit of 99\% confidence level for the semi-major axis of the
astrometric orbit. Unlike the significance analysis, our
estimated upper limits are not distribution-free.

The upper limit on the semi-major axis yields a lower limit for the
inclination and therefore an upper limit to the secondary mass for each
system. 

\section{Results}

\subsection{The Planet candidates}

As of March 2001, the Encyclopedia of extrasolar planets included 49
planet candidates with minimum masses smaller than 13 \MJ. In the
literature we found another 14 stars with secondaries with minimum
masses between 13 and 67 \MJ.  Although the separation between planets
and brown-dwarf secondaries is not yet clear (e.g., Mazeh and Zucker
2000) we nevertheless adopted the nomenclature of the
Encyclopedia and separated the discussion between the planet candidates
and the brown-dwarf secondaries. In this subsection we analyzed all but
two of the planet candidates. One of the 49 stars, BD $-$10\degr3166 had
no Hipparcos data, and therefore was not analyzed. The other star,
HD~168443, is known to have two companions.  We assume its reflex motion
is caused mainly by its heavier companion, whose mass is known to be
larger than 13\MJ. Therefore we defer its analysis to the brown-dwarf
subsection.

The Hipparcos and the spectroscopic data of the remaining 47 stars are
summarized in Table~1.  The table lists the stellar Hipparcos number,
name and distance; the eccentricity and period; the minimum size of the
astrometric semi-major axis and the minimum mass of the unseen
companion. The table then lists the number of Hipparcos data points, a
reference to the radial-velocity work and a comment column that
indicates, for example, whether the actual radial-velocity measurements
were available, or only the orbital elements.

In Table~2 we present the results of our analysis. After we list again
the Hipparcos number and stellar name,
we list $a_{derived}$, the derived semi-major axis, followed by its
uncertainty. We then
list $a_{99}$, derived from our permutation test. The next column
gives the 99\% upper limit of the astrometric orbit, $a_{upp-lim}$,
as derived by 
$$a_{upp-lim} =a_{derived}+2.3\sigma .$$ 
The corresponding 99\% upper limit for the mass is given in the last
column. Mass upper limits which are too large do
not contribute any additional information to our previous knowledge
about the companion. Therefore we decided, arbitrarily, to discard
upper limits larger than two solar masses.

Table 2 shows that all our $a_{derived}$ are smaller than
$a_{99}$. This includes the planets of $\upsilon$ And and HD10697
whose derived orbits were previously published (Mazeh et al.\ 1999;
Zucker \& Mazeh 2000), but the new analysis renders their orbits less significant. 
Figure~2, which depicts
$a_{derived}$ versus $a_{99}$, indeed shows that all points fall
below the line $a_{derived} = a_{99}$. This means that all our derived
astrometric motions are not significant in the level of 99\%.  However,
this does not mean that the orbits derived are all false.  
Figure 2 shows that some of the systems are close to the border line,
indicating that the orbits of these systems were detected with
significance close to 99\%. The systems with significance higher than
90\% are listed in Table~3. Here we list the Hipparcos number and the
stellar name, the confidence level of the derived astrometric orbit, the
derived semi-major axis, its uncertainty and the derived inclination;
the derived secondary mass, together with its 1$\sigma$ range. The
values in square brackets are the corresponding values calculated by
Pourbaix (2001), listed here for comparison.

\subsection{The Brown-Dwarf Candidates} 

In Table~4 and Table~5 we repeat the above analysis for the sub-stellar
candidates with minimum masses between 15 and 70 \MJ\ (i.e., brown-dwarf
candidates). 

Table 4 summarizes the Hipparcos and radial-velocity data of the
brown-dwarf candidates. The structure of the table is identical to that
of Table 1. The table includes HD~168443 which we ``expelled'' from Table
1. HD~98230 is a quadruple system and therefore its astrometric motion
must be quite complicated and thus was not analyzed. Nine of the
remaining 14 stars were analyzed by Halbwachs et al.\ (2000) and are marked
accordingly in the table. Table 5 presents the results of our analysis. 

Figure~3 depicts $a_{derived}$ versus $a_{99}$. Contrary to Figure~2,
 here a few systems are above the line $a_{derived} = a_{99}$,
 indicating significant detections.  As in the previous section, we list
 in Table~6 the systems for which our analysis indicated an astrometric
 orbit with a significance higher than 90\%. We indicate by an asterisk systems
 that were analyzed by Halbwachs et al.\ (2000). The structure of the
 table is similar to that of Table 3, except that the values in square
 brackets are the values obtained by Halbwachs et al., listed for
 comparison.

\section{HD~164427}
 
Table~6 includes one star --- HD~164427 --- that did not appear in the
Halbwachs et al.\ (2000) paper because its radial-velocity modulation
had not yet been detected.  We find that its derived astrometric orbit,
which renders its companion stellar, is significant on a 99\% confidence
level.  For HD~164427 we present in Figure~1b the histogram of the
falsely detected semi-major axes. Out of 1000 simulations only 11
yielded semi-major axis larger than 3.11 $mas$. It indicates that the
significance of this detection is 99\%, at about the 2.3$\sigma$ level.

Figure~4 presents the derived orbit of HD~164427. Traditionally, an
astrometric solution is presented graphically by a diagram of the
derived orbit on the plane of the sky, usually together with the
individual two-dimensional measurements. This is impossible for the
Hipparcos data, because these measurements are only one-dimensional,
observed along the instantaneous reference great circle at the time of
the measurement. However, two Hipparcos measurements observed at very
close timing with two different great-circle directions allow us, in
principle, to derive a two-dimensional stellar position. Because of the
intrinsic uncertainty of the measurements, more than two measurements
are desired for such an exercise. Such a ``two-dimensional averaging''
of the Hipparcos data was used by Halbwachs et al.\ (2000) as a graphic
representation of the astrometric orbit for the long-period binaries
they have studied. 

The graphical presentation of HD~164427 was done in a similar way.  We
folded the data with the orbital period and looked for small groups of
Hipparcos data points that cluster around the same orbital phase. Such
small clusters, with at least three points, were ``two-dimensionally
averaged''. The resulting points are presented in Figure~4a. The figure
shows that the points are at about 2--3 $mas$ away from the center, and
therefore indicates, although without any quantitative measure, that the
orbit is real.

Figure~4a does not show the temporal dependence of the points. To show
this dependence we derived the mean anomaly of each point from the
corresponding {\it observed} true anomaly, together with the other
orbital elements. In Figure~4b we plot the mean anomalies of the 8
points as a function of their orbital phase. Without noise, and
assuming there is a genuine orbital motion, we would
expect the points to lie along a straight line, which is also
plotted. The figure clearly demonstrates the nature of the stellar
orbital revolution. We stress again that these 8 points are used only
for graphic presentation and no real conclusion is drawn from them. Any
quantitative statement is based upon the full set of 28 one-dimensional
measurements.

\section{Discussion}

The analysis presented here shows that among the planet candidates no
orbit was detected with a significance higher than 99\%. Out of 47
systems, six orbits were derived with significance higher than 90\%.
These orbits are probably all false, as we expect to derive from the
whole sample $4.7 \pm 2.1$ false orbits with this significance or
higher. Three of these orbits were derived with significance higher
than 95\%, while we expect $2.3 \pm 1.5$ false orbits.  Although we
can not rule out the possibility that one or two of the six orbits are
real, apparently the Hipparcos precision is not enough to yield
detections of reflex motion induced by extrasolar planets with a high
enough statistical significance. This conclusion is contrary to the
preliminary suggestion of \HBGs, who found at least four astrometric
orbits highly significant, and agrees well with the general conclusion
of Pourbaix (2001) and Pourbaix \& Arenou (2001).

Although there is no difference in our conclusion, it is interesting
to compare in details our results with those of Pourbaix (2001).
Pourbaix found one case --- HD~195019 --- where the astrometric fit is
improved by the orbital model at the 99\% level.  We also find the
orbit of HD~195019 to be somewhat significant, but to a much smaller
degree --- 92\%. We suspect that the difference in the results could
have come from the fact that Pourbaix used strong assumptions about
the distribution of the errors of the Hipparcos data, while we do
not. In any case, both Pourbaix and Pourbaix \& Arenou (2001)
suggested that the astrometric orbit is probably not real.

Another difference, which is again of academic interest, is the case
of HD~6434.  Here the difference in the significance assigned by the
two analyses is so large that we suspect that Pourbaix analysis
converged to a completely different orbit, with substantial smaller
significance. A well-known phenomenon with the Hipparcos data is the
existence of two minima of the $\chi^2$ function corresponding to two
almost opposite orientations of the orbit, where one minimum is deeper
than the other and thus corresponds to a better fit. Our analysis
finds Pourbaix's solution for this star at the shallower minimum.  In
any case, both analyses suggest that the Hipparcos orbit of this
system is not real.

We turn now to discuss the upper limits we derived for the planet
candidates.  The smallest value listed in Table~2 is that of 47~UMa,
with an upper limit of 0.014 \Mo. This upper limit is at about the
value commonly adopted {\it arbitrarily} as the border line between
planets and brown dwarfs. This means that our analysis indicates that
47 UMa companion can almost safely be regarded as a planet. Perryman
et al.\ (1996) found an upper limit of 22~\MJ\ for a confidence level
of 90\%. If we follow their prescription for the 99\% limit we get
28~\MJ. This is twice as large as our upper limit, probably because
Perryman et al.\ did not use all the information available from the
radial-velocity solution for their astrometric fit. The same
difference can be found in the analysis of 70 Vir. We find a 99\%
upper limit of 63~\MJ, while Perryman et al.\ found a limit of 85 \MJ.

For eight stars the upper limit is below 40 \MJ, and for another six
stars the upper limit is below 80 \MJ.  These findings negate the
preliminary conjecture (\HBGs) that many of the planet candidates are
disguised stellar companions with extremely small angles.

On the other hand, our analysis of the brown-dwarf candidates yielded
quite a few astrometric orbits with high significance. Out of 14 systems
we found 8 orbits with significance higher than 90\%, while we expected
to find only $1.4 \pm 1.1$ false orbits under the assumption that no
real orbits are present.
This suggests that most of them are real. Similarly,
we find six orbits with significance higher than 95\%, while we expected
only $0.7 \pm 0.8$ false orbits. This again suggested that most of these
orbits are real. Further support to this suggestion can be
found in the fact that all the corresponding orbital inclinations,
except one, are large, as opposed to the ones of Table~3.

This is not a surprise. The stellar reflex motion caused by a brown
dwarf is much larger than the one caused by a planet. Therefore the
minimum semi-major axes, $a\times \sin i$, of the brown-dwarf candidates
are closer to the threshold of the Hipparcos detection. This is
reflected, for example, by the fact that out of the 14 brown-dwarf
candidate systems only two systems, with short orbital periods, have $a\times
\sin i$ smaller than 0.1 $mas$, while the planets have 34 such orbits
out of 47 systems. Only one of these two brown-dwarf systems, HD~87330,
was found to have an astrometric orbit with a significance higher than
90\%. Here again the inclination is suspiciously very small. We therefore
suspect that this is a false orbit. After all we expect $1.4
\pm 1.1$ systems to show such false orbits.

All the significant orbits of the brown-dwarf candidates yielded
secondary masses that rendered the secondary a stellar object. This fact
was pointed out already by Halbwachs et al.\ (2000), who analyzed six of
the eight systems of Table~6, and got very similar results for the
secondary masses.  Halbwachs et al.\ (2000) analyzed another five of the
brown-dwarf candidates, which we find their orbit
insignificant. Although Halbwachs et al.\ (2000) did not give explicitly
the significance of their finding, their derived large uncertainty and
the following discussion leave no doubt that these orbits are
insignificant. So, we agree on these five systems as well. Note that these
five systems include HD~114762, which we still considered a planet {\it
candidate}. In the sample we analyzed we still had six
systems that could still be brown-dwarf secondaries. 

To summarize, we find no disguised low-mass stellar companions within
the sample of 47 stars that harbor planet companions, while we find
6--8 such stellar companions in a sample of 14 stars with brown-dwarf
candidates. The frequency of unknown low-mass stellar companions
detected as planet candidates depends on the selection of the sample
from which the planets are searched. Apparently, the planet hunters avoided any
known spectroscopic binaries in their sample, relying on previous lower
precision radial-velocity searches. Obviously, some M-star secondaries
could have avoided previous detection. However, such binaries must have
extremely small inclination angles in order to be erroneously
identified as planets. This is why we do not find any such case. 

On the other hand, even moderate inclinations could turn binaries with
stellar low-mass secondaries into brown-dwarf candidates.  This is why we
find relatively many stellar secondaries in the brown-dwarf candidate
sample. Maybe there are no brown-dwarfs at all, and the brown-dwarf
candidates are all low-mass stellar secondaries.  The fact that we found
no compelling evidence for a disguised brown-dwarf secondary within the
planet candidates is consistent with the ``brown-dwarf desert''
conjecture.  However, as pointed out by Halbwachs et al.\ (2000), the
brown-dwarf desert conjecture is not proved by these findings.  It is still
possible that the other 6--8 systems in Table~4 have brown-dwarf
companions. In addition, our derived upper limits of the planet
candidates can rule out brown-dwarf candidates only for a very few
systems. and therefore the real number of hidden brown dwarfs within the
sample of planet candidates is still not known.

The possible brown-dwarf desert separates the two certainly inhabited
lands of the planets on one side and that of the stellar secondaries on
the other side, and therefore is crucial to our understanding of the two
populations. To find out more about it we need a more detailed analysis of
the mass distribution of the two populations, a study which is underway.

\acknowledgments

We are indebted to Yoav Benjamini for illuminating
discussions with regard to the derivation of upper limits.  This work
was supported by the US-Israel Binational Science Foundation through
grant 97-00460 and the Israeli Science Foundation (grant no. 40/00)

\newpage

\begin{figure}
\plotone{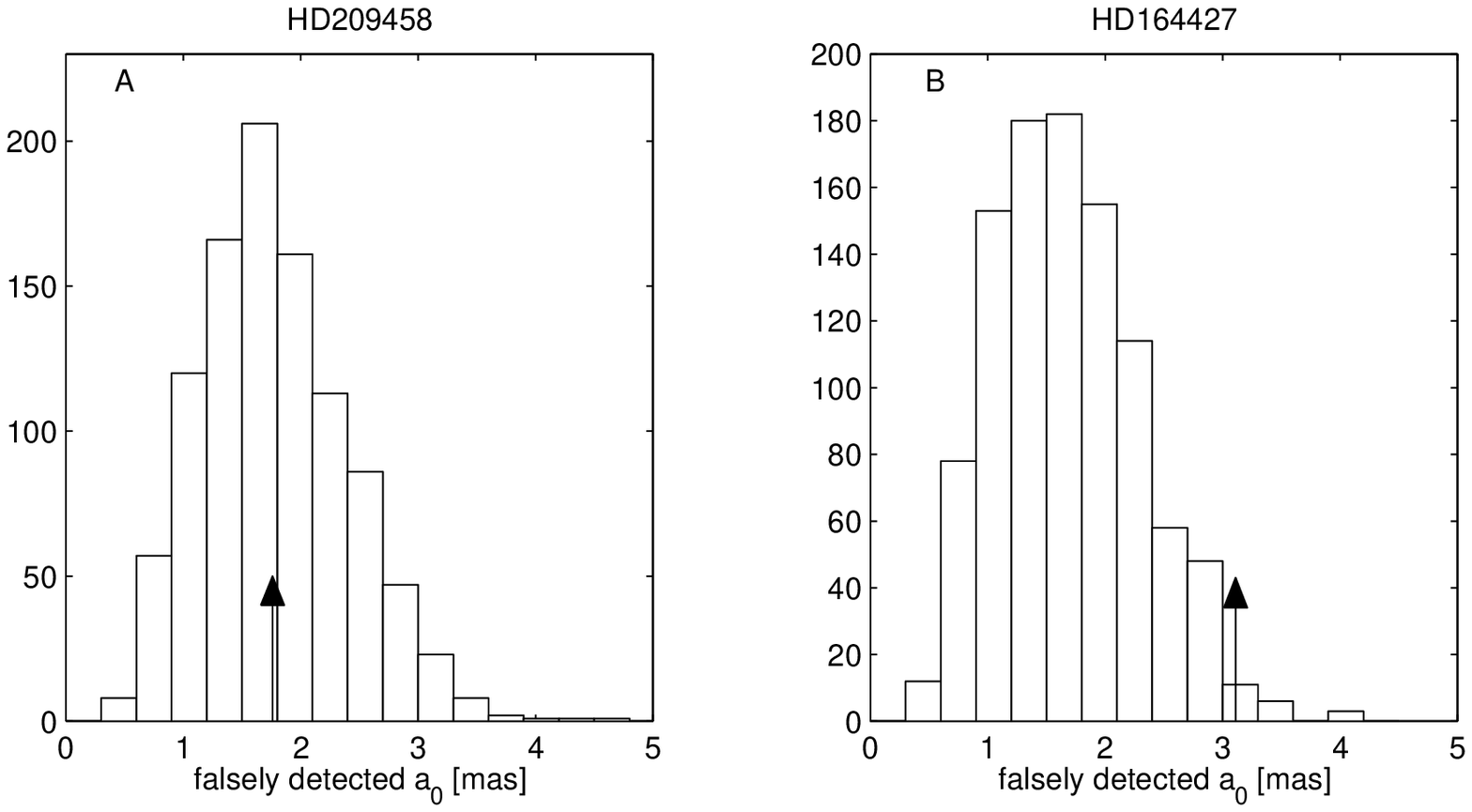}
\caption{Histograms of the size of the falsely
  derived semi-major
axes in the simulated permuted data. The size of the actually detected
axis is marked by an arrow. \label{fig1}}
\end{figure}

\newpage

\begin{figure}
\plotone{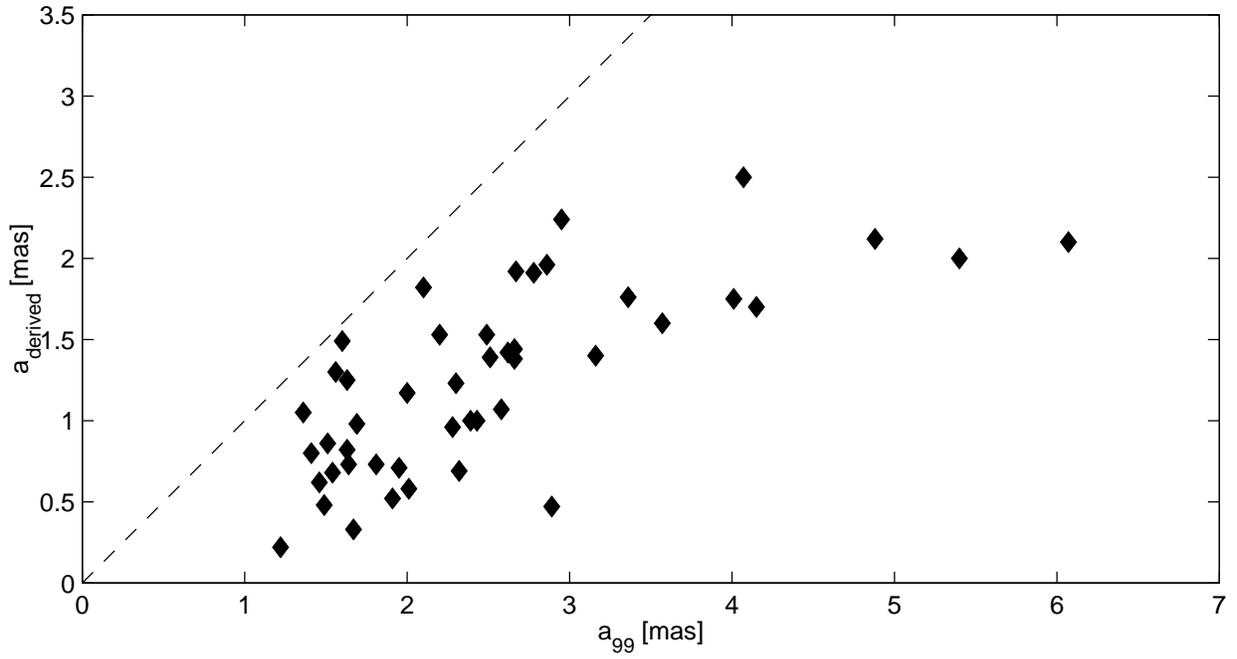}
\caption{A plot of the derived semi-major axis
  of the planet candidates as a
  function of the 99th percentile of the falsely derived
  semi-major axes. The dashed line represents the line $a_{derived} =
  a_{99}$. The star $\epsilon$ Eri was excluded from the plot because of
  its peculiarly high $a_{99}$. \label{fig2}}
\end{figure}

\newpage

\begin{figure}
\plotone{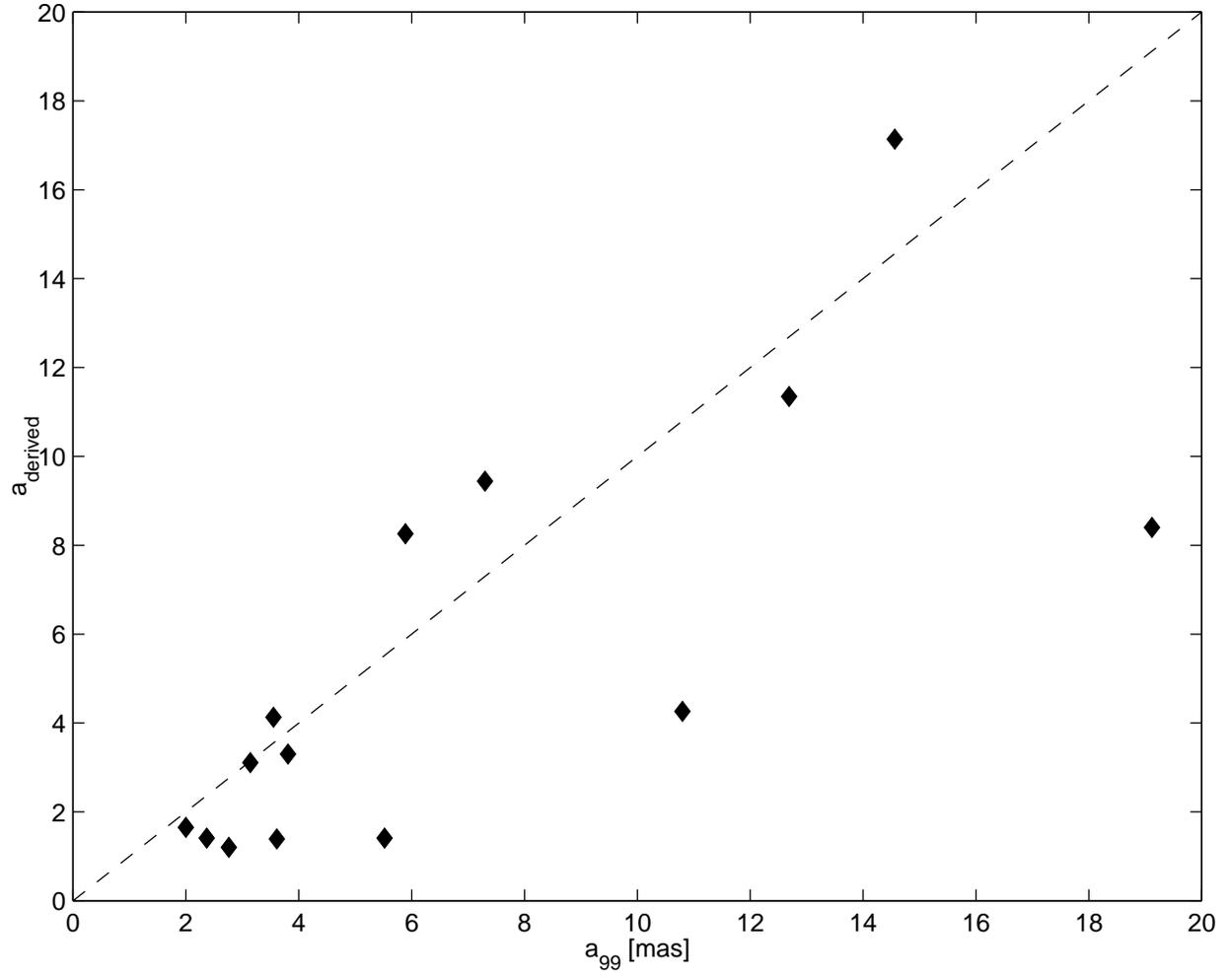}
\caption{A plot of the derived semi-major axis
  of the brown-dwarf candidates as a
  function of the 99th percentile of the falsely derived
  semi-major axes. The dashed line represents the line $a_{derived} =
  a_{99}$. \label{fig3}}
\end{figure}

\newpage

\begin{figure}
\plotone{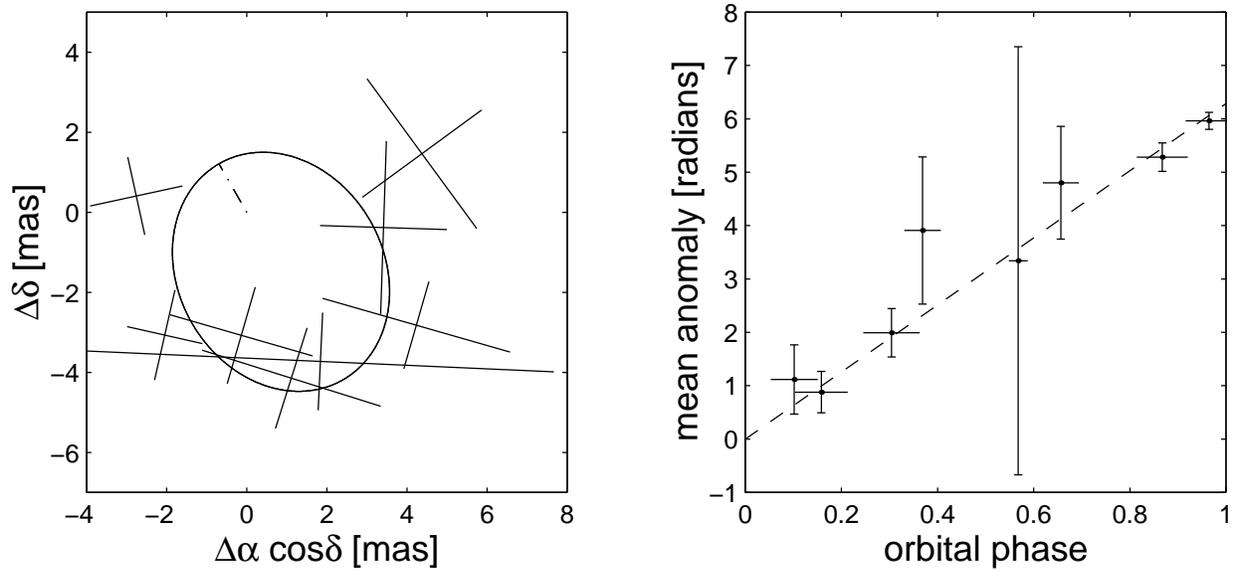}
\caption{A. Eight ``two-dimensionally averaged''
  Hipparcos data points of HD~164427.
The derived astrometric orbit is also plotted. The
direction of the periastron is marked by a dot-dashed line. 
B. The mean anomaly of the 8 points as
a function of their orbital phase. The dashed line indicates the
expected dependency for points without noise. \label{fig4}}
\end{figure}

\begin{deluxetable}{rlclcllcrc}
\tabletypesize{\footnotesize}
\tablewidth{0pt}
\tablecolumns{10}
\tablecaption{Planet candidates list. \label{table1}}
\tablehead{
\colhead{HIP} & 
\colhead{Name} & 
\colhead{$d$} & 
\colhead{$e$} & 
\colhead{$P$} &
\colhead{$a\sin i$} & 
\colhead{$M\sin i$} & 
\colhead{$N_F$/$N_N$/$N_O$} & 
\colhead{References} & 
\colhead{Commments} \\
\colhead{number} &
\colhead{} &
\colhead{(pc)} &
\colhead{} &
\colhead{(days)} &
\colhead{($mas$)} &
\colhead{(\MJ)} &
\colhead{} &
\colhead{} &
\colhead{}
}
\startdata
1292   & GJ 3021        & 18  & 0.51 & 133.82 & 0.099  & 3.3   & 41/42/43 & 1  & a   \\
5054   & HD 6434        & 40  & 0.30 &  22.09 & 0.0018 & 0.48  & 40/39/40 & 2  & a   \\
7513   & $\upsilon$ And & 13  & 0.41 & 1267   & 0.58   & 4.6   & 26/28/28 & 3  & a,b \\
8159   & HD 10697       & 33  & 0.12 & 1072.3 & 0.35   & 6.3   & 16/17/17 & 4  & c   \\
9683   & HD 12661       & 37  & 0.23 & 252.7  & 0.055  & 2.8   & 26/26/26 & 5  & d   \\
10138  & Gliese 86      & 11  & 0.05 & 15.78  & 0.050  & 4     & 34/34/35 & 6  & a   \\
12048  & HD 16141       & 36  & 0.28 & 75.82  & 0.0020 & 0.21  & 16/17/17 & 7  & c   \\
12653  & HR 810         & 17  & 0.16 & 320.1  & 0.10   & 2.3   & 38/38/39 & 8  & c   \\
14954  & HD 19994       & 22  & 0.2  & 454    & 0.082  & 2.0   & 26/26/27 & 2  & a,e \\
16537  & $\epsilon$ Eri & 3.2 & 0.61 & 2502.1 & 1.08   & 0.86  & 34/34/18 & 9  & a,f \\
19921  & HD 27442       & 18  & 0.03 & 426.5  & 0.074  & 1.4   & 31/36/36 & 10 & c   \\
26381  & HD 37124       & 33  & 0.19 & 155.7  & 0.018  & 1.0   & 18/17/18 & 4  & c   \\
27253  & HD 38529       & 42  & 0.27 & 14.32  & 0.0016 & 0.81  & 19/19/38 & 5  & a,g \\
31246  & HD 46375       & 33  & 0.   & 3.02   & 0.0003 & 0.25  & 18/19/20 & 7  & c   \\
33719  & HD 52265       & 28  & 0.29 & 118.96 & 0.017  & 1.1   & 24/25/25 & 11 & c   \\
43177  & HD 75289       & 29  & 0.   & 3.51   & 0.0006 & 0.42  & 42/43/44 & 12 & a   \\
43587  & 55 Cnc         & 13  & 0.05 & 14.65  & 0.0083 & 0.84  & 24/25/25 & 13 & a   \\
47007  & HD 82943       & 27  & 0.61 & 442.6  & 0.086  & 2.2   & 25/25/26 & 14 & a   \\
47202  & HD 83443       & 44  & 0.42 & 29.83  & 0.0008 & 0.16  & 36/36/38 & 15 & a,b \\
50786  & HD 89744       & 39  & 0.7  & 256.   & 0.11   & 7.2   & 24/27/27 & 16 & a   \\
52409  & HD 92788       & 32  & 0.30 & 326.7  & 0.10   & 3.3   & 18/22/22 & 5  & d   \\
53721  & 47 UMa         & 14  & 0.03 & 1090   & 0.32   & 2.4   & 28/30/30 & 17 & a   \\
60644  & HD 108147      & 39  & 0.56 & 10.88  & 0.0008 & 0.34  & 33/35/36 & 14 & a   \\
64426  & HD 114762      & 41  & 0.33 & 84.03  & 0.11   & 11    & 18/18/20 & 18 & c   \\
65721  & 70 Vir         & 18  & 0.4  & 116.7  & 0.17   & 6.6   & 29/32/33 & 19 & a   \\
67275  & $\tau$ Boo     & 16  & 0.02 & 3.31   & 0.0092 & 3.9   & 27/27/27 & 13 & a   \\
68162  & HD 121504      & 44  & 0.13 & 64.6   & 0.0060 & 0.89  & 32/37/38 & 2  & a   \\
72339  & HD 130322      & 30  & 0.05 & 10.72  & 0.0038 & 1.0   & 15/13/15 & 12 & a   \\
74500  & HD 134987      & 26  & 0.24 & 259.6  & 0.044  & 1.6   & 17/17/17 & 4  & c   \\
78459  & $\rho$ CrB     & 17  & 0.03 & 39.65  & 0.014  & 1.1   & 41/43/43 & 20 & c   \\
79248  & 14 Her         & 18  & 0.32 & 1654   & 0.58   & 3.3   & 39/41/41 & 14 & a   \\
79336  & HD 187123      & 48  & 0.   & 3.10   & 0.0004 & 0.48  & 41/42/43 & 4  & c   \\
86796  & HD 160691      & 15  & 0.62 & 743    & 0.19   & 2.0   & 31/32/32 & 10 & c   \\
90004  & HD 168746      & 43  & 0.   & 6.41   & 0.0004 & 0.24  & 18/23/23 & 14 & a   \\
90485  & HD 169830      & 36  & 0.34 & 230.4  & 0.045  & 3.0   & 21/22/22 & 14 & a   \\
93746  & HD 177830      & 59  & 0.41 & 391.6  & 0.018  & 1.2   & 48/50/50 & 4  & c   \\
94645  & HD 179949      & 27  & 0.   & 3.09   & 0.0012 & 0.84  & 20/22/23 & 21 & c   \\
96901  & 16 Cyg B       & 21  & 0.63 & 801    & 0.12   & 1.5   & 37/39/39 & 22 & d   \\
98714  & HD 190228      & 62  & 0.5  & 1161   & 0.14   & 5     & 56/58/58 & 23 & a   \\
99711  & HD 192263      & 20  & 0.22 & 24.36  & 0.0074 & 0.79  & 24/25/25 & 4  & c   \\
100970 & HD 195019      & 37  & 0.02 & 18.2   & 0.013  & 3.5   & 23/23/24 & 4  & c   \\
108859 & HD 209458      & 47  & 0.   & 3.52   & 0.0006 & 0.69  & 24/26/26 & 24 & a,h \\
109378 & HD 210277      & 21  & 0.45 & 437    & 0.068  & 1.3   & 25/25/26 & 25 & c   \\
113020 & Gliese 876     & 4.7 & 0.34 & 60.97  & 0.28   & 2.0   & 17/19/19 & 26 & a   \\
113357 & 51 Peg         & 15  & 0.   & 4.23   & 0.0014 & 0.47  & 36/34/37 & 27 & c   \\
113421 & HD 217107      & 20  & 0.14 & 7.13   & 0.0047 & 1.3   & 20/21/21 & 4  & a,g \\
116906 & HD 222582      & 42  & 0.71 & 575.9  & 0.17   & 5.3   & 21/24/24 & 4  & c   \\
\enddata
\tablecomments{
$N_F$, $N_N$ and $N_O$ are the numbers of the Hipparcos FAST measurements, NDAC measurements and 
the number of satellite orbits included in the star's analysis,
respectively. The comments: \\
a. The analysis used the published orbital elements. \\
b. The analysis pertains to the outer planet of a multiple planets
system. \\
c. The analysis used the available radial velocities. \\
d. The analysis used two independent sets of radial velocities. \\
e. The analysis assumed the planet orbit is circular. \\
f. The orbital period is much longer than the Hipparcos mission
duration. \\
g. The Hipparcos analysis used an additional acceleration term. \\
h. The planet is known to eclipse the star.
}
\tablerefs{ 
(1) Naef et al.\ 2000; 
(2) Queloz et al.\ 2000b; 
(3) Laughlin \& Adams 1999; 
(4) Vogt et al.\ 2000; 
(5) Fischer et al.\ 2000; 
(6) Queloz et al.\ 2000a; 
(7) Marcy, Butler \& Vogt 2000; 
(8) Kurster et al.\ 2000; 
(9) Hatzes et al.\ 2000; 
(10) Butler et al.\ 20001; 
(11) Butler et al.\ 2000; 
(12) Udry et al.\ 2000a; 
(13) Butler et al.\ 1997; 
(14) Udry et al.\ 2001;
(15) Mayor et al.\ 2000; 
(16) Korzennik et al.\ 2000; 
(17) Butler \& Marcy 1996; 
(18) Latham, private communication;
(19) Marcy \& Butler 1996; 
(20) Noyeset al.\ 1997; 
(21) Tinney et al.\ 2001; 
(22) Cochran et al.\ 1997; 
(23) Sivan et al.\ 2000; 
(24) Mazeh et al.\ 2000; 
(25) Marcy et al.\ 1999;
(26) Delfosse et al.\ 1998; 
(27) Marcy et al.\ 1997
}
\end{deluxetable}

\begin{deluxetable}{rlllrrl}
\tabletypesize{\scriptsize}
\tablewidth{0pt}
\tablecolumns{7}
\tablecaption{Results of the Planet candidates analysis. \label{table2}}
\tablehead{
\colhead{HIP} & 
\colhead{Name} & 
\colhead{$a_{derived}$} & 
\colhead{$\sigma_{a}$} & 
\colhead{$a_{99}$} &
\colhead{$a_{upp-lim}$} & 
\colhead{$M_{upp-lim}$} \\
\colhead{number} &
\colhead{} &
\colhead{($mas$)} &
\colhead{($mas$)} &
\colhead{($mas$)} &
\colhead{($mas$)} &
\colhead{(\Mo)} 
}
\startdata
1292   &  GJ 3021        & 0.48 & 0.50 &  1.49  & 1.62  & 0.054   \\
5054   &  HD 6434        & 1.3  & 0.67 &  1.56  & 2.88  & 1.31    \\
7513   &  $\upsilon$ And & 1.42 & 0.68 &  2.62  & 2.98  & 0.018   \\
8159   &  HD 10697       & 2.12 & 0.73 &  4.88  & 3.80  & 0.088   \\
9683   &  HD 12661       & 0.96 & 0.64 &  2.28  & 2.43  & 0.13    \\
10138  &  Gliese 86      & 0.33 & 0.74 &  1.67  & 2.03  & 0.14    \\
12048  &  HD 16141       & 1.0  & 1.2  &  2.39  & 3.70  & 0.51    \\
12653  &  HR 810         & 0.80 & 0.56 &  1.41  & 2.09  & 0.042   \\
14954  &  HD 19994       & 0.98 & 0.72 &  1.69  & 2.63  & 0.063   \\
16537  &  $\epsilon$ Eri & 10.1 & 7.0  & 26.49  & 26.21 & 0.021   \\
19921  &  HD 27442       & 0.82 & 0.47 &  1.63  & 1.90  & 0.036   \\
26381  &  HD 37124       & 2.1  & 1.8  &  6.07  & 6.22  & 0.46    \\
27253  &  HD 38529       & 1.7  & 1.1  &  4.15  & 4.36  & \nodata \\
31246  &  HD 46375       & 1.4  & 1.0  &  3.16  & 3.71  & \nodata \\
33719  &  HD 52265       & 0.62 & 0.56 &  1.46  & 1.91  & 0.13    \\
43177  &  HD 75289       & 1.05 & 0.52 &  1.36  & 2.24  & \nodata \\
43587  &  55 Cnc         & 0.58 & 0.80 &  2.01  & 2.41  & 0.28    \\
47007  &  HD 82943       & 1.39 & 0.88 &  2.51  & 3.42  & 0.090   \\
47202  &  HD 83443       & 1.96 & 0.68 &  2.86  & 3.52  & 1.40    \\
50786  &  HD 89744       & 1.44 & 0.87 &  2.66  & 3.44  & 0.18    \\
52409  &  HD 92788       & 1.53 & 0.79 &  2.49  & 3.34  & 0.094   \\
53721  &  47 UMa         & 0.47 & 0.68 &  2.89  & 2.04  & 0.014   \\
60644  &  HD 108147      & 0.71 & 0.67 &  1.95  & 2.26  & 1.80    \\
64426  &  HD 114762      & 1.07 & 0.93 &  2.58  & 3.21  & 0.27    \\
65721  &  70 Vir         & 0.73 & 0.50 &  1.64  & 1.88  & 0.063   \\
67275  &  $\tau$ Boo     & 0.52 & 0.56 &  1.91  & 1.81  & 1.23    \\
68162  &  HD 121504      & 1.91 & 0.81 &  2.78  & 3.77  & 0.77    \\
72339  &  HD 130322      & 2.5  & 1.5  &  4.07  & 5.92  & \nodata \\
74500  &  HD 134987      & 1.0  & 1.1  &  2.43  & 3.49  & 0.13    \\
78459  &  $\rho$ CrB     & 1.49 & 0.44 &  1.60  & 2.51  & 0.22    \\
79248  &  14 Her         & 1.38 & 0.83 &  2.66  & 3.29  & 0.019   \\
79336  &  HD 187123      & 0.22 & 0.57 &  1.22  & 1.54  & \nodata \\
86796  &  HD 160691      & 1.17 & 0.67 &  2.00  & 2.71  & 0.028   \\
90004  &  HD 168746      & 1.53 & 0.84 &  2.20  & 3.47  & \nodata \\
90485  &  HD 169830      & 1.25 & 0.64 &  1.63  & 2.71  & 0.18    \\
93746  &  HD 177830      & 0.86 & 0.49 &  1.51  & 1.98  & 0.13    \\
94645  &  HD 179949      & 1.92 & 0.68 &  2.67  & 3.49  & \nodata \\
96901  &  16 Cyg B       & 0.73 & 0.57 &  1.81  & 2.05  & 0.026   \\
98714  &  HD 190228      & 1.82 & 0.77 &  2.10  & 3.59  & 0.074   \\
99711  &  HD 192263      & 1.6  & 1.1  &  3.57  & 4.19  & 0.63    \\
100970 &  HD 195019      & 2.24 & 0.78 &  2.95  & 4.03  & 1.41    \\
108859 &  HD 209458      & 1.76 & 0.90 &  3.36  & 3.84  & \nodata \\
109378 &  HD 210277      & 0.69 & 0.61 &  2.32  & 2.08  & 0.038   \\
113020 &  Gliese 876     & 2.0  & 1.9  &  5.40  & 6.41  & 0.049   \\
113357 &  51 Peg         & 0.68 & 0.63 &  1.54  & 2.13  & 1.04    \\
113421 &  HD 217107      & 1.23 & 0.85 &  2.30  & 3.18  & 1.60    \\
116906 &  HD 222582      & 1.75 & 0.94 &  4.01  & 3.91  & 0.14    \\
\enddata
\end{deluxetable}

\begin{deluxetable}{rllllllc}
\tabletypesize{\scriptsize}
\tablewidth{0pt}
\tablecolumns{9}
\tablecaption{Derived orbits of planet candidates with a confidence
  level higher than 90\%. \label{table3}}
\tablehead{
\colhead{HIP} & 
\colhead{Name} & 
\colhead{p} & 
\colhead{$a_{derived}$} & 
\colhead{$\sigma_{a}$} &
\colhead{$i_{derived}$} & 
\colhead{$M_{derived}$} & 
\colhead{Mass Range} \\
\colhead{number} &
\colhead{} &
\colhead{} &
\colhead{($mas$)} &
\colhead{($mas$)} &
\colhead{(deg)} &
\colhead{(\Mo)} &
\colhead{(1$\sigma$)} 
}
\startdata
5054   & HD 6434    & 0.96 [0.49]    & 1.34 & 0.67 & -0.08 [0.2] & 0.45  & (0.20,0.77)   \\
43177  & HD 75289   & 0.90           & 1.05 & 0.52 & 0.03        & 1.13  & (0.45,2.19)   \\
78459  & $\rho$ CrB & 0.98 [0.99]    & 1.49 & 0.46 & 0.54 [-0.9] & 0.12  & (0.086,0.17)  \\
90485  & HD 169830  & 0.92 [0.90]    & 1.25 & 0.64 & 2.1 [2.1]   & 0.081 & (0.039,0.124) \\
94645  & HD 179949  & 0.90           & 1.92 & 0.68 & 0.034       & 3.4   & (1.57,6.49)   \\
98714  & HD 190228  & 0.95 [0.91]    & 1.82 & 0.77 & 4.5 [5.1]   & 0.064 & (0.037,0.093) \\
100970 & HD 195019  & 0.92 [$>$0.99] & 2.24 & 0.78 & 0.32 [0.3]  & 0.92  & (0.51,1.47)   \\
\enddata
\end{deluxetable}

\begin{deluxetable}{rlclcllcrc}
\tabletypesize{\scriptsize}
\tablewidth{0pt}
\tablecolumns{10}
\tablecaption{Brown dwarf candidates list. \label{table4}}
\tablehead{
\colhead{HIP} & 
\colhead{Name} & 
\colhead{$d$} & 
\colhead{$e$} & 
\colhead{$P$} &
\colhead{$a\sin i$} & 
\colhead{$M\sin i$} & 
\colhead{$N_F$/$N_N$/$N_O$} & 
\colhead{References} & 
\colhead{Commments} \\
\colhead{number} &
\colhead{} &
\colhead{(pc)} &
\colhead{} &
\colhead{(days)} &
\colhead{($mas$)} &
\colhead{(\MJ)} &
\colhead{} &
\colhead{} &
\colhead{}
}
\startdata
$^\ast$13769  & HD 18445   & 26 & 0.56 & 555    & 2.6178  & 44 & 26/29/29 & 1 & a   \\
$^\ast$19832  & BD $-$04\degr782 & 20 & 0.07 & 717    & 3.7883  & 47 & 27/26/27 & 1 & a   \\
$^\ast$21482  & HD 283750  & 18 & 0.   & 1.788  & 0.0966  & 50 & 22/22/22 & 1 & a   \\
$^\ast$21832  & HD 29587   & 28 & 0.36 & 1470   & 4.0125  & 41 & 11/12/12 & 2 & b   \\
$^\ast$50671  & HD 89707   & 35 & 0.95 & 297.71 & 1.2924  & 57 & 25/26/26 & 1 & a   \\
$^\ast$62145  & HD 110833  & 15 & 0.78 & 271.2  & 1.1159  & 17 & 45/47/47 & 1 & a   \\
$^\ast$63366  & HD 112758  & 21 & 0.14 & 103.26 & 0.7434  & 33 & 22/23/24 & 1 & a   \\
$^\ast$70950  & HD 127506  & 22 & 0.72 & 2599   & 6.8006  & 36 & 53/55/56 & 1 & a   \\
$^\ast$77152  & HD 140913  & 48 & 0.61 & 147.96 & 0.4618  & 46 & 54/57/57 & 1 & a   \\
87330         & HD 162020  & 31 & 0.28 & 8.43   & 0.0418  & 14 & 16/18/18 & 3 & c   \\
88531         & HD 164427  & 39 & 0.55 & 108.55 & 0.4603  & 46 & 24/28/28 & 4 & b   \\
89844         & HD 168443  & 38 & 0.27 & 1667   & 1.1024  & 15 & 15/18/19 & 5 & c,d \\
104903        & HD 202206  & 46 & 0.42 & 259    & 0.2542  & 15 & 36/35/36 & 3 & c   \\
$^\ast$113718 & HD 217580  & 17 & 0.52 & 454.7  & 4.8412  & 67 & 18/17/19 & 1 & a,e \\
\enddata
\tablecomments{$N_F$, $N_N$ and $N_O$ are the numbers of the Hipparcos
  FAST measurements, NDAC measurements and the number of satellite orbits 
  included in the star's analysis. The asterisks mark stars that were
  already analyzed by Halbwachs et al. The comments: \\
a. The analysis used the published orbital elements, where $a\sin i$
was available instead of the radial velocity amplitude. \\
b. The analysis used the available radial velocities. \\
c. The analysis used the published orbital elements. \\
d. The analysis pertains to the outer companion. \\
e. A similar solution already appears in the Hipparcos catalog.
}
\tablerefs{ 
(1) Halbwachs et al.\ 2000;
(2) Latham, private communication;
(3) http://obswww.unige.ch/~udry/planet; 
(4) Tinney et al.\ 2001; 
(5) Udry et al.\ 2000b;
}
\end{deluxetable}

\begin{deluxetable}{rlclrrl}
\tabletypesize{\scriptsize}
\tablewidth{0pt}
\tablecolumns{7}
\tablecaption{Results of the brown dwarf candidates analysis. \label{table5}}
\tablehead{
\colhead{HIP} & 
\colhead{Name} & 
\colhead{$a_{derived}$} & 
\colhead{$\sigma_{a}$} & 
\colhead{$a_{99}$} &
\colhead{$a_{upp-lim}$} & 
\colhead{$M_{upp-lim}$} \\
\colhead{number} &
\colhead{} &
\colhead{($mas$)} &
\colhead{($mas$)} &
\colhead{($mas$)} &
\colhead{($mas$)} &
\colhead{(\Mo)} 
}
\startdata
$^\ast$13769  & HD 18445   & 9.44  & 0.95 & 7.30  & 9.66  & 0.13     \\
$^\ast$19832  & BD $-$04\degr782 & 17.14 & 0.84 & 14.56 & 19.07 & 0.20     \\
$^\ast$21482  & HD 283750  & 1.20  & 0.77 & 2.76  & 2.97  & \nodata  \\
$^\ast$21832  & HD 29587   & 4.26  & 0.96 & 10.80 & 6.47  & 0.13     \\
$^\ast$50671  & HD 89707   & 1.39  & 0.58 & 3.61  & 2.72  & 0.16     \\
$^\ast$62145  & HD 110833  & 8.26  & 0.61 & 5.89  & 9.66  & 0.093    \\
$^\ast$63366  & HD 112758  & 4.13  & 0.72 & 3.55  & 5.79  & 0.17     \\
$^\ast$70950  & HD 127506  & 8.4   & 3.9  & 19.12 & 17.37 & 0.10     \\
$^\ast$77152  & HD 140913  & 1.65  & 0.63 & 2.00  & 3.10  & 0.21     \\
87330         & HD 162020  & 3.3   & 1.5  & 3.81  & 6.75  & \nodata  \\
88531         & HD 164427  & 3.11  & 0.66 & 3.14  & 4.63  & 0.36     \\ 
89844         & HD 168443  & 1.41  & 0.82 & 5.52  & 3.30  & 0.077    \\
104903        & HD 202206  & 1.41  & 0.87 & 2.37  & 3.41  & 0.14     \\
$^\ast$113718 & HD 217580  & 11.35 & 0.82 & 12.69 & 13.24 & 0.18     \\
\enddata
\end{deluxetable}

\begin{deluxetable}{rllllclc}
\tabletypesize{\scriptsize}
\tablewidth{0pt}
\tablecolumns{9}
\tablecaption{Derived orbits of brown-dwarf candidates with a
  confidence level higher than 90\%. \label{table6}}
\tablehead{
\colhead{HIP} & 
\colhead{Name} & 
\colhead{p} & 
\colhead{$a_{derived}$} & 
\colhead{$\sigma_{a}$} &
\colhead{$i_{derived}$} & 
\colhead{$M_{derived}$} & 
\colhead{Mass Range} \\
\colhead{number} &
\colhead{} &
\colhead{} &
\colhead{($mas$)} &
\colhead{($mas$)} &
\colhead{(deg)} &
\colhead{(\Mo)} &
\colhead{(1$\sigma$)}
}
\startdata
$^\ast$13769  & HD 18445   & 0.998    & 9.44 [9.85]   & 0.95 &  -16.1 & 0.167 & (0.148,0.186) \\
$^\ast$19832  & BD $-$04\degr782 & 0.999    & 17.14 [17.15] & 0.84 &  12.77 & 0.242 & (0.227,0.256) \\
$^\ast$62145  & HD 110833  & $>$0.999 & 8.26 [8.39]   & 0.61 &  7.76  & 0.134 & (0.123,0.145) \\
$^\ast$63366  & HD 112758  & 0.996    & 4.13 [4.13]   & 0.72 &  10.4  & 0.20  & (0.16,0.24)   \\
$^\ast$77152  & HD 140913  & 0.92     & 1.65 [1.64]   & 0.63 &  16.3  & 0.17  & (0.10,0.24)   \\
87330         & HD 162020  & 0.94     & 3.3           & 1.5  &  0.73  & 3.30  & (1.04,8.08)   \\
88531         & HD 164427  & 0.99     & 3.11          & 0.66 &  8.5   & 0.35  & (0.27,0.45)   \\
$^\ast$113718 & HD 217580  & 0.97     & 11.35 [11.31] & 0.82 &  25.2  & 0.162 & (0.148,0.175) \\
\enddata
\end{deluxetable}

\end{document}